\documentstyle[vsolj01,graphicx,natbib]{article}

\RequirePackage[T1]{fontenc}

\def\cite{\citealt}

\begin{document}

\title{V606 Aql (Nova Aquilae 1899) is now a dwarf nova}

\author{Taichi Kato$^1$ and Naoto Kojiguchi$^1$}
\author{$^1$ Department of Astronomy, Kyoto University,
       Sakyo-ku, Kyoto 606-8502, Japan}
\email{tkato@kusastro.kyoto-u.ac.jp}

\begin{abstract}
We found that the 1899 nova V606 Aql currently shows
dwarf nova outbursts with a typical cycle length of
270~d and amplitudes of $\sim$1.5 mag using
Public Data Release of Zwicky Transient Facility
observations.  The low mass-transfer rate in quiescence
has been suggested to explain the large eruption amplitude
\citep{tap16oldnova}, and the present detection of dwarf nova
outbursts supports this interpretation.  The transition
to the dwarf nova state more than 100~yr after the nova
eruption gives credence to the hibernation scenario.
The absolute magnitude estimated from dwarf nova outbursts
suggests that V606 Aql should have been a fast nova
and the presence of high excitation lines in quiescence
would be explained by the presence of a massive white dwarf.
\end{abstract}

   V606 Aql is a nova discovered by W. Fleming.
The nova was first seen on a plate on 1899 April 21.
The early stage of the nova eruption was not well covered
by observations.  The brightest magnitude recorded was
6.7 (pg).  The light curve showed a steep decline,
followed by a plateau of 100~d duration
\citep{due87novaatlas}.  \citet{due87novaatlas} extrapolated
the light curve and estimated the peak of 5.5 and
$t_3$ of 65~d.

   The identification of the post-nova remained
a mystery for a long period despite the bright
magnitude during the nova eruption.
\citet{kha71novaID} proposed an identification of
a blend of stars.  \citet{due87novaatlas} listed
three stars composing this blend.  The exact identification,
however, was finally made by \citet{tap16oldnova}
by multicolor photometry and spectroscopy.

   The post-nova had brightness of $V$=20.4 and $R$=20.1
\citep{tap16oldnova}.  The amplitude of the nova eruption
was one of the largest among the sample of \citet{tap16oldnova}
and there were only $\sim$10 novae with similar amplitudes
in the sample of pre-1986 eruptions \citep{tap16oldnova}.
\citet{tap16oldnova} discussed that these large amplitudes
of eruptions may be associated with the low mass-transfer
rates ($\dot{M}$) in the quiescent disk.

\begin{figure*}
  \begin{center}
    \includegraphics[width=16cm]{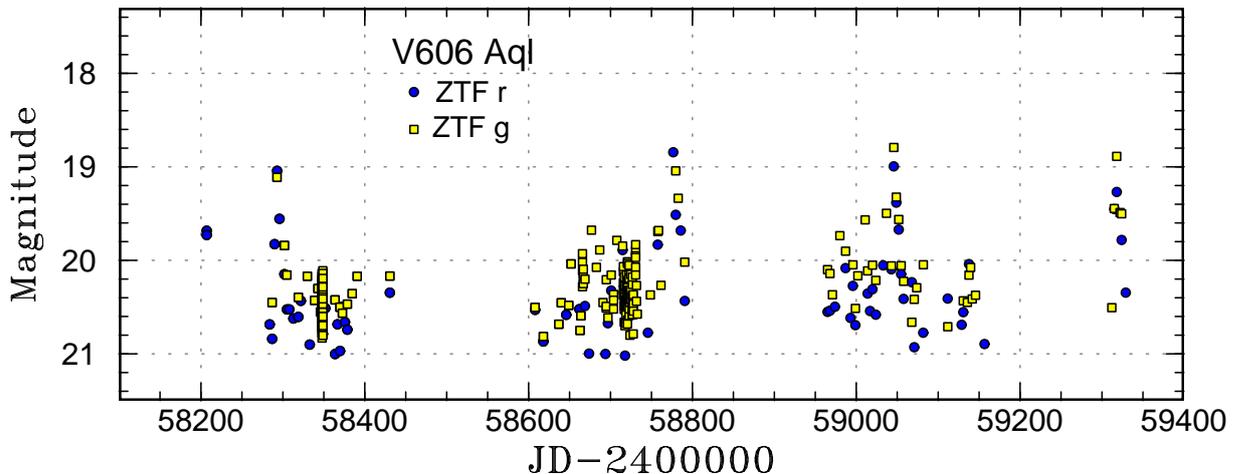}
  \end{center}
  \caption{ZTF light curve of V606 Aql}
  \label{fig:v606aql}
\end{figure*}

   We used Public Data Release 6 of
the Zwicky Transient Facility (ZTF, \cite{ZTF})
observations\footnote{
   The ZTF data can be obtained from IRSA
$<$https://irsa.ipac.caltech.edu/Missions/ztf.html$>$
using the interface
$<$https://irsa.ipac.caltech.edu/docs/program\_interface/ztf\_api.html$>$
or using a wrapper of the above IRSA API
$<$https://github.com/MickaelRigault/ztfquery$>$.
} and found that V606 Aql is currently in dwarf nova state
(figure \ref{fig:v606aql}).

   There were four dwarf nova outbursts between 2018 June
and 2021 April (table \ref{tab:outbursts}).  Note that ZTF
$g$ and $r$ observations were not simultaneous.
The intervals of the second, third and the fourth outbursts
were 269~d and 272~d.  The interval between the first
and second outbursts was 484~d, suggesting that one
outburst was missed between them near the solar conjunction.
The relatively stable intervals between the outbursts
also support the dwarf nova-type instability.
The outbursts lasted for $\sim$10~d and the fading rates
were around 0.1 mag~d$^{-1}$.  Although the fading rates
suggest and object with a relatively long orbital period
[i.e. more than several hours, see e.g. \citet{war87CVabsmag};
\citet{war95book}], these fading rates were likely severely
underestimated due to the contribution from
the bright quiescent disk.  The amplitudes of the outbursts
were about 1.5~mag.
ZTF made time-resolved observations on two nights on
2018 August 17 and 18 (less than 3~hr each) without
detecting an eclipse.

   The dwarf nova state is what is expected for
a post-nova with a reduced $\dot{M}$, as suggested by
\citet{tap16oldnova}.  This finding is in line with
the ``hibernation'' scenario (\cite{Hibernation};
\cite{pat84CVevolution}), which suggests that $\dot{M}$
in post-novae is reduced long time ($\sim$100~yr or more)
after the nova eruption.  The most recent detection
of the dwarf nova activity was in the old nova BC Cas (1929)
\citep{kat20bccas}.  These recent findings give credence to 
the hibernation scenario.

   The typical absolute magnitude of outbursting
dwarf novae is $+$4 \citep{war87CVabsmag} with a dependence
on the orbital period.  Since the orbital period of
V606 Aql is unknown, we used $+$4.
Using the estimated $E(B-V)$=0.35 \citep{tap16oldnova}
and the standard $A(V)/E(B-V)$ = 3.1, the distance
modulus is estimated to be 13.8.
The observed brightest magnitude of the 1899
nova eruption corresponds to $M_{\rm pg}=-8.5$
(corrected for extinction).
Using the Gaia-calibrated maximum magnitude - 
rate of decline (MMRD) relation \citep{ozd18novaMMRD},
this value corresponds to a fast nova with
$t_3$ of $\sim$15~d.  The value appears to be consistent
with the ``steep decline'' of the nova eruption.
The presence of a plateau phase, however, may have been
unusual for a fast nova.
This discussion needs to be treated with caution since 
the true maximum was not observed and since it is unclear whether
the entire disk becomes hot during these dwarf nova
outbursts in post-novae.

   The presence of He II and Bowen emission
lines \citep{tap16oldnova} is not typical for
a dwarf nova.  Considering the large eruption amplitude
and steep decline of the nova
outburst, the mass of the white dwarf is expected
to be large (e.g. \cite{hac06novadecline}).
The existence of these high excitation lines may be
a result of a massive white dwarf as in \citet{kat21bocet}.

\begin{table*}
\caption{List of dwarf nova outbursts}\label{tab:outbursts}
\begin{center}
\begin{tabular}{cccc}
\hline
Peak date & JD & ZTF & ZTF \\
          &    & $g$ & $r$ \\
\hline
2018 June 23    & 2458293 & 19.04 & 19.11 \\
2019 October 20 & 2458777 & 18.84 & -- \\
2020 July 15    & 2459046 & 19.03 & 18.85 \\
2021 April 13   & 2459318 & 19.27 & 18.89 \\
\hline
\end{tabular}
\end{center}
\end{table*}

\section*{Acknowledgments}

This work was supported by JSPS KAKENHI Grant Number 21K03616.

Based on observations obtained with the Samuel Oschin 48-inch
Telescope at the Palomar Observatory as part of
the Zwicky Transient Facility project. ZTF is supported by
the National Science Foundation under Grant No. AST-1440341
and a collaboration including Caltech, IPAC, 
the Weizmann Institute for Science, the Oskar Klein Center
at Stockholm University, the University of Maryland,
the University of Washington, Deutsches Elektronen-Synchrotron
and Humboldt University, Los Alamos National Laboratories, 
the TANGO Consortium of Taiwan, the University of 
Wisconsin at Milwaukee, and Lawrence Berkeley National Laboratories.
Operations are conducted by COO, IPAC, and UW.

The ztfquery code was funded by the European Research Council
(ERC) under the European Union's Horizon 2020 research and 
innovation programme (grant agreement n$^{\circ}$759194
-- USNAC, PI: Rigault).

\end{document}